\documentstyle[12pt,epsf]{article}
\begin{document}
\tolerance=5000
\def\pp{{\, \mid \hskip -1.5mm =}}
\def\cL{{\cal L}}
\def\be{\begin{equation}}
\def\ee{\end{equation}}
\def\bea{\begin{eqnarray}}
\def\eea{\end{eqnarray}}
\def\tr{{\rm tr}\, }
\def\nn{\nonumber \\}
\def\e{{\rm e}}
\def\D{{D \hskip -3mm /\,}}

\def\SEH{S_{\rm EH}}
\def\SGH{S_{\rm GH}}
\def\AdS5{{{\rm AdS}_5}}
\def\S4{{{\rm S}_4}}
\def\gfv{{g_{(5)}}}
\def\gfr{{g_{(4)}}}
\def\SC{{S_{\rm C}}}
\def\RH{{R_{\rm H}}}


\  \hfill 
\begin{minipage}{3.5cm}
NDA-FP-74 \\
May 2000 \\
\end{minipage}

\vfill

\begin{center}
{\large\bf (Non)-singular brane-world cosmology induced by quantum effects 
in d5 dilatonic gravity}

\vfill

{\sc Shin'ichi NOJIRI}\footnote{email: nojiri@cc.nda.ac.jp}, 
{\sc Octavio OBREGON}$^{\spadesuit}$\footnote{
octavio@ifug3.ugto.mx}, \\
and {\sc Sergei D. ODINTSOV}$^{\spadesuit}$\footnote{
On leave from Tomsk State Pedagogical University, RUSSIA. \\
\ \hskip 1cm email: odintsov@ifug5.ugto.mx}, \\

\vfill

{\sl Department of Applied Physics \\
National Defence Academy, 
Hashirimizu Yokosuka 239, JAPAN}

\vfill

{\sl $\spadesuit$ 
Instituto de Fisica de la Universidad de 
Guanajuato \\
Apdo.Postal E-143, 37150 Leon, Gto., MEXICO}

\vfill

{\bf ABSTRACT}

\end{center}

5d dilatonic gravity (bosonic sector of gauged supergravity) with 
non-trivial bulk potential and with surface terms (boundary 
cosmological constant and trace anomaly induced effective action 
for brane quantum matter) is considered. For constant bulk 
potential and maximally SUSY Yang-Mills theory (CFT living 
on the brane) the inflationary brane-world is constructed. 
The bulk is singular asymptotically  AdS space with non-constant 
dilaton and dilatonic de Sitter or hyperbolic brane is induced 
by quantum matter effects. At the same time, dilaton on the brane 
is determined dynamically. This all is natural realization of 
warped compactification in AdS/CFT correspondence. For fine-tuned toy 
example of non-constant bulk potential we found the non-singular 
dilatonic brane-world where bulk again represents asymptotically 
AdS space and de Sitter brane (inflationary phase of observable 
Universe) is induced exclusively by quantum effects. The radius 
of the brane and dilaton are 
determined dynamically. The analytically solvable example 
of exponential bulk potential leading to singular asymptotically AdS 
dilatonic bulk space with de Sitter (or hyperbolic) brane is also presented.
In all cases under discussion the gravity 
on the brane is trapped via Randall-Sundrum scenario. It is shown 
that qualitatively the same types of brane-worlds occur when 
quantum brane matter is described by $N$ dilaton coupled spinors. 

\noindent
PACS number(s): 04.70.-s

\newpage

\section{Introduction}

After the discovery that gravity on the brane may be localized \cite{RS}
there was renewed interest in the studies of higher-dimensional 
(brane-world) theories. In particular, numerous works \cite{CH} (and 
refs. therein) have been devoted to the investigation of cosmology 
(inflation) of brane-worlds. In refs.\cite{NOZ,HHR,inf} it has been 
suggested the 
inflationary brane-world scenario realized due to quantum effects of 
brane matter. Such scenario is based on large $N$ quantum CFT living 
on the brane \cite{NOZ,HHR}. Actually, that corresponds to 
implementing of RS compactification 
within the context of renormalization group flow in AdS/CFT set-up. 
Note that working within large $N$ approximation justifies 
such approach to brane-world quantum cosmology as then 
quantum matter loops contribution is essential.

Another important aspect of brane-world Universe with localized 
gravity is related with the possibility to resolve the 
cosmological constant problem. For example, it has been 
shown in ref.\cite{ADKS} that in the presence of bulk scalar 
(dilaton) one can find  static solutions 
of equations of motion where the bulk dilatonic potential vanishes. Such 
self-tuning mechanism has been further studied in ref.\cite{Youm}.
Unfortunately, it is usual  that such solutions which localize 
gravity have a naked space-time singularity. (The presence of 
non-trivial dilatonic potential makes the situation even 
more complicated \cite{GJS}). General 
properties of 
the self-tuning domain wall solutions of 5d gravity-scalar system 
with various potentials and brane couplings have been discussed in 
ref.\cite{CEGH}.
It has been shown there that for some specific potential 
the resolution of singularities (when potential and brane 
coupling are fine-tuned) may be 
achieved. The bulk spacetime is asymptotically AdS and gravity 
localization may occur without having singularities!
However, in the studies in this direction the discussion has been 
done so far  mainly for 
solutions with flat 4d domain walls (flat branes). (The explicit, 
non-singular example of ref.\cite{CEGH} 
corresponds to such flat brane configuration).
The reason is that in this case the second-order equations of motion
may be reduced to first-order form for an arbitrary dilatonic 
potential, see ref.\cite{Pot} for explicit examples. 
In the case when the branes 
are not flat this procedure does not work directly, 
generally speaking.
\footnote{Note that brane-world theory may be often understood as 
completely 4d two measure theory\cite{ADDK}. Such models with exponential 
potentials which appear due to scale invariance have been intensively
studied in refs.\cite{Guendelman}(inflation and role of vacuum effects).}

The purpose of the present work is to investigate the role
of quantum matter living on the brane in the study of brane-world 
cosmology in 5d AdS dilatonic gravity with non-trivial dilatonic 
potential (bosonic sector of the corresponding gauged supergravity). 
We are mainly interested in 
the situation when the boundary of 5d AdS space represents 
 a 4d constant curvature space whose creation (as is shown) 
is possible only due to
quantum effects of brane matter.
Thus, the possibility of dilatonic brane-world inflation induced
by quantum effects is proved. In different versions of such 
scenario discussed here 
the dynamical determination of dilaton occurs as well.

The paper is organized as follows. In the next section we investigate 
dilatonic brane-world inflation induced by quantum effects (using
 anomaly induced effective action) in the situation
with constant bulk potential. The bulk space represents (singular) 
asymptotically AdS background with non-trivial dilaton. 
The brane matter quantum effects (maximally SUSY Yang-Mills 
theory is considered as brane CFT) help to create de Sitter or 
AdS space on the brane. Hence, the quantum realization of 
brane-world inflation is possible in the presence of the dilaton 
which is determined dynamically in the bulk as well as on the brane.
Note that an analytical treatment is done in this section. 
Section 3 is devoted to extension of results of previous 
consideration for non-constant bulk potentials. 
One solvable example of bulk 
equations of motion for exponential potential is given. 
In this case de Sitter (or hyperbolic) brane with small radius occurs for 
SUSY Yang-Mills theory. It is also interesting that without quantum 
corrections ($W$ vanishes) the dilatonic hyperbolic brane is 
still possible. 

The conditions to 
get non-singular, asymptotically AdS dilatonic spacetime (when 4d gravity
 is trapped) are 
discussed. An example
of a toy dilatonic potential is presented. It is shown (in some 
approximation) that due to 
quantum effects the brane represents de Sitter space 
and localization of gravity occurs. Hence, the role of 
such (fine-tuned) dilatonic potential is to make weaker 
(or to avoid completely) the singularity which appears for the  
AdS bulk solution of section 2. Dilaton is still determined 
dynamically.
In section 4 we show that all the above picture may be well realized 
in the situation when brane matter is not exactly conformal 
invariant matter 
(dilaton coupled spinors are considered). Similar qualitative 
results as in previous sections are obtained. 
Some resume and perspectives are drawn in the Discussion. 
In the Appendix a short discussion of some equivalence between 
5d dilatonic gravity and 4d dilatonic gravity coupled 
with CFT is done.

\section{\label{I}Dilatonic brane-world inflation induced by quantum 
effects: Constant bulk potential}

We start with Euclidean signature 
for the action $S$ which is the sum of 
the Einstein-Hilbert action $\SEH$ with kinetic term for dilaton $\phi$, 
the Gibbons-Hawking surface term $\SGH$,  the surface 
counter term $S_1$ and the trace anomaly induced action 
$W$\footnote{For the introduction to anomaly induced 
effective action in curved space-time (with torsion), see
section 5.5 in \cite{BOS}.}: 
\bea
\label{Stotal}
S&=&\SEH + \SGH + 2 S_1 + W, \\
\label{SEHi}
\SEH&=&{1 \over 16\pi G}\int d^5 x \sqrt{\gfv}\left(R_{(5)} 
 -{1 \over 2}\nabla_\mu\phi\nabla^\mu \phi 
 + {12 \over l^2}\right), \\
\label{GHi}
\SGH&=&{1 \over 8\pi G}\int d^4 x \sqrt{\gfr}\nabla_\mu n^\mu, \\
\label{S1}
S_1&=& -{3 \over 8\pi G l}\int d^4 x \sqrt{\gfr}, \\
\label{W}
W&=& b \int d^4x \sqrt{\widetilde g}\widetilde F A \nn
&& + b' \int d^4x\left\{A \left[2{\widetilde\Box}^2 
+\widetilde R_{\mu\nu}\widetilde\nabla_\mu\widetilde\nabla_\nu 
 - {4 \over 3}\widetilde R \widetilde\Box^2 
+ {2 \over 3}(\widetilde\nabla^\mu \widetilde R)\widetilde\nabla_\mu
\right]A \right. \nn
&& \left. + \left(\widetilde G - {2 \over 3}\widetilde\Box \widetilde R
\right)A \right\} \nn
&& -{1 \over 12}\left\{b''+ {2 \over 3}(b + b')\right\}
\int d^4x \left[ \widetilde R - 6\widetilde\Box A 
 - 6 (\widetilde\nabla_\mu A)(\widetilde \nabla^\mu A)
\right]^2 \nn
&& + C \int d^4x 
A \phi \left[{\widetilde\Box}^2 
+ 2\widetilde R_{\mu\nu}\widetilde\nabla_\mu\widetilde\nabla_\nu 
 - {2 \over 3}\widetilde R \widetilde\Box^2 
+ {1 \over 3}(\widetilde\nabla^\mu \widetilde R)\widetilde\nabla_\mu
\right]\phi \ .
\eea 
Here the quantities in the  5 dimensional bulk spacetime are 
specified by the suffices $_{(5)}$ and those in the boundary 4 
dimensional spacetime are specified by $_{(4)}$. 
The factor $2$ in front of $S_1$ in (\ref{Stotal}) is coming from 
that we have two bulk regions which 
are connected with each other by the brane. 
In (\ref{GHi}), $n^\mu$ is 
the unit vector normal to the boundary. In (\ref{W}), one chooses 
the 4 dimensional boundary metric as 
\be
\label{tildeg}
\gfr_{\mu\nu}=\e^{2A}\tilde g_{\mu\nu},
\ee 
and we specify the 
quantities given by $\tilde g_{\mu\nu}$ by using $\tilde{\ }$. 
$G$ ($\tilde G$) and $F$ ($\tilde F$) are the Gauss-Bonnet
invariant and the square of the Weyl tensor, which are given as
\bea
\label{GF}
G&=&R^2 -4 R_{ij}R^{ij}
+ R_{ijkl}R^{ijkl}, \nn
F&=&{1 \over 3}R^2 -2 R_{ij}R^{ij}
+ R_{ijkl}R^{ijkl} \ ,
\eea
\footnote{We use the following curvature conventions:
\begin{eqnarray*}
R&=&g^{\mu\nu}R_{\mu\nu} \nn
R_{\mu\nu}&=& R^\lambda_{\ \mu\lambda\nu} \nn
R^\lambda_{\ \mu\rho\nu}&=&
-\Gamma^\lambda_{\mu\rho,\nu}
+ \Gamma^\lambda_{\mu\nu,\rho}
- \Gamma^\eta_{\mu\rho}\Gamma^\lambda_{\nu\eta}
+ \Gamma^\eta_{\mu\nu}\Gamma^\lambda_{\rho\eta} \nn
\Gamma^\eta_{\mu\lambda}&=&{1 \over 2}g^{\eta\nu}\left(
g_{\mu\nu,\lambda} + g_{\lambda\nu,\mu} - g_{\mu\lambda,\nu} 
\right)\ .
\end{eqnarray*}}
In the effective action (\ref{W}), 
we now consider the case corresponding to ${\cal N}=4$ 
$SU(N)$ Yang-Mills theory, where \cite{LT}
\be
\label{bbC}
b=-b'={C \over 4}
={N^2 -1 \over 4(4\pi)^2}\ .
\ee
The dilaton field $\phi$ which appears from the coupling 
with extended conformal supergravity is in general  
complex but we consider the case in which that only the real part of 
$\phi$ is 
non-zero. Adopting AdS/CFT correspondence one can argue that in symmetric 
phase the quantum brane matter appears due to maximally SUSY Yang-Mills 
theory as above. Note that there is a kinetic term for the dilaton 
in the classical bulk action but also there is dilatonic contribution to 
the anomaly induced effective action $W$. Here, it appears the difference 
with the correspondent construction in ref.\cite{inf} where there was no 
dilaton.  

In the bulk, the solution of the equations of motion 
is given in \cite{NOtwo}, as follows
\bea
\label{curv2}
ds^2&=&f(y)dy^2 + y\sum_{i,j=0}^{d-1}\hat g_{ij}(x^k)dx^i dx^j \nn
f&=&{d(d-1)  \over 4y^2
\lambda^2 \left(1 + { c^2 \over 2\lambda^2 y^d}
+ {kd \over \lambda^2 y}\right)} \nn
\phi&=&c\int dy \sqrt{{d(d-1) \over
4y^{d +2}\lambda^2 \left(1 + { c^2 \over 2\lambda^2 y^d}
+ {kd \over \lambda^2 y}\right)}}\ .
\eea
Here $\lambda^2={12 \over l^2}$ and 
$\hat g_{ij}$ is the metric of the Einstein manifold, which is
defined by $r_{ij}=k\hat g_{ij}$, where $r_{ij}$ is 
the Ricci tensor constructed with $\hat g_{ij}$ and 
$k$ is a constant. 
We should note that there is a curvature singularity at
$y=0$ \cite{NOtwo}. 
The solution with non-trivial dilaton would presumbly correspond to 
the deformation of the vacuum (which is associated 
with the dimension 4 operator, say ${\rm tr} F^2$) in the dual 
maximally SUSY Yang-Mills theory. 

If one defines a new coordinate $z$ by
\be
\label{c2b}
z=\int dy\sqrt{d(d-1)  \over 4y^2
\lambda^2 \left(1 + { c^2 \over 2\lambda^2 y^d}
+ {kd \over \lambda^2 y}\right)},
\ee
and solves $y$ with respect to $z$, we obtain the warp
factor $\e^{2\hat A(z,k)}=y(z)$. Here one assumes 
the metric of 5 dimensional space time as follows:
\be
\label{metric1}
ds^2=dz^2 + \e^{2A(z,\sigma)}\tilde g_{\mu\nu}dx^\mu dx^\nu\ ,
\quad \tilde g_{\mu\nu}dx^\mu dx^\nu\equiv l^2\left(d \sigma^2 
+ d\Omega^2_3\right)\ .
\ee
Here $d\Omega^2_3$ corresponds to the metric of 3 dimensional 
unit sphere. Then for the unit sphere ($k=3$), we find
\be
\label{smetric}
A(z,\sigma)=\hat A(z,k=3) - \ln\cosh\sigma\ ,
\ee
for the flat Euclidean space ($k=0$)
\be
\label{emetric}
A(z,\sigma)=\hat A(z,k=0) + \sigma\ ,
\ee
and for the unit hyperboloid ($k=-3$)
\be
\label{hmetric}
A(z,\sigma)=\hat A(z,k=-3) - \ln\sinh\sigma\ .
\ee
We now identify $A$ and $\tilde g$ in (\ref{metric1}) with those in 
(\ref{tildeg}). Then we find $\tilde F=\tilde G=0$, 
$\tilde R={6 \over l^2}$ etc. 

According to the assumption in (\ref{metric1}), the actions in (\ref{SEHi}), 
(\ref{GHi}), (\ref{S1}), and (\ref{W}) have the following forms:
\bea
\label{SEHii}
\SEH&=& {l^4 V_3 \over 16\pi G}\int dz d\sigma \left\{\left( -8 
\partial_z^2 A - 20 (\partial_z A)^2\right)\e^{4A} \right. \nn
&& +\left(-6\partial_\sigma^2 A - 6 (\partial_\sigma A)^2 
+ 6 \right)\e^{2A} \nn
&& \left. -{1 \over 2}\e^{4A}(\partial_z\phi)^2
-{1 \over 2l^2}\e^{2A}(\partial_\sigma\phi)^2
+ {12 \over l^2} \e^{4A}\right\}, \\
\label{GHii}
\SGH&=& {3l^4 V_3 \over 8\pi G}\int d\sigma \e^{4A} 
\partial_z A, \\
\label{S1ii}
S_1&=& - {3l^3 V_3 \over 8\pi G}\int d\sigma \e^{4A}, \\
\label{Wii}
W&=& V_3 \int d\sigma \left[b'A\left(2\partial_\sigma^4 A
 - 8 \partial_\sigma^2 A \right) \right. \nn
&& - 2(b + b')\left(1 - \partial_\sigma^2 A 
 - (\partial_\sigma A)^2 \right)^2 \nn
&&\left. +CA\phi\left(\partial_\sigma^4 \phi
 - 4 \partial_\sigma^2 \phi \right) \right]\ .
\eea
Here $V_3$ is the volume or area of the unit 3 sphere: 
\be
\label{Csmr7}
V_3=2\pi^2\ .
\ee

On the brane at the boundary, 
one gets the following equations 
\bea
\label{eq2}
0&=&{48 l^4 \over 16\pi G}\left(\partial_z A - {1 \over l}
\right)\e^{4A}
+b'\left(4\partial_\sigma^4 A - 16 \partial_\sigma^2 A\right) \nn
&& - 4(b+b')\left(\partial_\sigma^4 A + 2 \partial_\sigma^2 A 
 - 6 (\partial_\sigma A)^2\partial_\sigma^2 A \right) \nn
&& + 2C\left(\partial_\sigma^4 \phi
 - 4 \partial_\sigma^2 \phi \right), 
\eea
from the variation over $A$ and 
\be
\label{eq2p}
0=-{l^4 \over 8\pi G}\e^{4A}\partial_z\phi
+ C\left\{A\left(\partial_\sigma^4 \phi
 - 4 \partial_\sigma^2 \phi \right) 
+ \partial_\sigma^4 (A\phi)
 - 4 \partial_\sigma^2 (A\phi) \right\},
\ee 
from the variation over $\phi$. 
We should note that the contributions from $\SEH$ and $\SGH$ are 
twice from the naive values since we have two bulk regions which 
are connected with each other by the brane. 
The equations (\ref{eq2}) and (\ref{eq2p}) do not depend on 
$k$, that is, they are correct for any of the sphere, 
hyperboloid, or flat Euclidean space. The $k$ dependence appears 
when the bulk solutions are substituted. 
Substituting the bulk solution given by (\ref{curv2}), 
(\ref{c2b}) and (\ref{smetric}), (\ref{emetric}) or 
(\ref{hmetric}) into (\ref{eq2}) and (\ref{eq2p}), one obtains
\bea
\label{slbr1}
0&=&{1 \over \pi G l}\left(\sqrt{1 + {kl^2 \over 3y_0} 
+ {l^2 c^2 \over 24 y_0^4}} 
- 1\right)y_0^2 + 8 b', \\
\label{slbr1p}
0&=& -{c \over 8\pi G}+ 6C\phi_0 \ .
\eea
Here we assume the brane lies at $y=y_0$ and the dilaton takes 
a constant value there $\phi=\phi_0$: 
\be
\label{phi0}
\phi_0={c \over 48\pi G C}\ .
\ee
Note that eq.(\ref{slbr1}) does not depend on $b$ and $C$. 
Eq.(\ref{slbr1p}) determines the value of $\phi_0$. That might 
be interesting since the vacuum expectation value of the dilaton 
cannot be determined perturbatively in  string theory. 
Of course, (\ref{phi0}) contains the parameter $c$, which indicates   
 the non-triviality of the dilaton. 
The parameter $c$, however, can be determined from (\ref{slbr1}). 
Hence, in such scenario one gets a dynamical mechanism to determine 
 of dilaton on the boundary (in our observable world).

The effective tension of the domain wall is given by 
\be
\label{tF}
\sigma_{\rm eff}={3 \over 4\pi G }\partial_y A
={3 \over 4\pi G l}\sqrt{1 + {kl^2 \over 3y_0} 
+ {l^2 c^2 \over 24 y_0^4}}\ .
\ee
We should note that the radial ($z$) component of the geodesic equation  
 for the  in the metric (\ref{metric1}) 
is given by ${d^2x^z \over d\tau^2} + \partial_z A \e^{2A}
\left({dx^t \over d\tau}\right)^2=0$. Here $\tau$ is the 
proper time and we can normalize $\e^{2A}
\left({dx^t \over d\tau}\right)^2 = 1$ and obtain 
${d^2x^z \over d\tau^2} + \partial_z A=0$. Since the cosmological 
constant on the brane is given by ${3 \over 4\pi G }$, 
$\sigma_{\rm eff}$ gives the effective mass density: 
${3 \over 4\pi G }{d^2x^z \over d\tau^2} = - \sigma_{\rm eff}$.

As in \cite{HHR}, defining the radius $R$ of the brane in the 
following way
\be
\label{R}
R^2\equiv y_0\ ,
\ee
we can rewrite (\ref{slbr1}) as 
\be
\label{slbr2}
0={1 \over \pi G l}\left(\sqrt{1 + {kl^2 \over 3R^2} 
+ {l^2 c^2 \over 24 R^8}} -1 \right)R^4 + 8 b' \ .
\ee
Especially when the dilaton vanishes ($c=0$) and the brane is 
the unit sphere ($k=3$), the equation (\ref{slbr2}) reproduces the 
result of ref.\cite{HHR} for ${\cal N}=4$ $SU(N)$ super Yang-Mills 
theory in case of the large $N$ limit where 
$b'\rightarrow -{N^2 \over 4(4\pi )^2}$: 
\be
\label{slbr3}
{R^3 \over l^3}\sqrt{1 + {R^2 \over l^2}}={R^4 \over l^4}
+ {GN^2 \over 8\pi l^3}\ .
\ee 

Let us define a function $F(R, c)$ as 
\be
\label{FRc}
F(R,c)\equiv {1 \over \pi G l}\left(\sqrt{1 + {kl^2 \over 3R^2} 
+ {l^2 c^2 \over 24 R^8}} -1 \right)R^4 \ ,
\ee
It appears in the r.h.s. in (\ref{slbr2}). 

First we consider the $k>0$ case. Since 
\bea
\label{FRc2}
{\partial \left(\ln F(R,c)\right) \over \partial R}
&=&{1 \over R}\left(\sqrt{1 + {kl^2 \over 3R^2} 
+ {l^2 c^2 \over 24 R^8}} -1 \right)^{-1}
\left(\sqrt{1 + {kl^2 \over 3R^2} 
+ {l^2 c^2 \over 24 R^8}} \right)^{-1} \nn
&& \times \left(4 + {kl^2 \over R^2} 
+ 4\sqrt{1 + {kl^2 \over 3R^2} 
+ {l^2 c^2 \over 24 R^8}} \right)^{-1} \nn
&& \times \left({8kl^2 \over 3R^2} + {k^2 l^4 \over R^4}
 - {2l^2 c^2 \over 3 R^8}\right) \ .
\eea
$F(R,c)$ has a minimum at $R=R_0$, where $R_0$ is defined by
\be
\label{min}
0={8kl^2 \over 3R_0^2} + {k^2 l^4 \over R_0^4}
 - {2l^2 c^2 \over 3 R_0^8}\ .
\ee
When $k>0$, there is only one solution for $R_0$. 
Therefore $F(R,c)$ in the case of $k>0$ (sphere case) 
is a monotonically increasing function of $R$ when 
$R>R_0$ and a decreasing function when $R<R_0$. 
Since $F(R,c)$ is clearly a monotonically increasing 
function of $c$, we find for $k>0$ and $b'<0$ case 
that $R$ decreases when $c$ increases if $R>R_0$, that is, 
the non-trivial dilaton makes the radius smaller. 
Then, since $1/R$ corresponds to the rate of the 
inflation of the universe, when we Wick-rotate the sphere 
into the inflationary universe, the large dilaton supports the 
rapid universe expansion. 
Hence, we showed that quantum CFT living on the domain wall leads 
to the creation of inflationary dilatonic 4d de Sitter-brane Universe 
realized within 5d AdS bulk space.\footnote{Such brane-world quantum 
inflation for the case of constant dilaton has been presented in 
refs.\cite{NOZ,HHR,inf}. In the usual 4d world the anomaly induced inflation 
has been suggested in ref.\cite{SMM} 
(no dilaton) and in ref.\cite{Brevik} 
when a non-constant dilaton is present.} Of 
course, such ever expanding 
inflationary brane-world is understood in a sense of the analytical 
continuation of 4d sphere to Lorentzian signature. It would be 
interesting to understand the relation between such inflationary 
brane-world and inflation in D-branes, for example, of Hagedorn type
 \cite{AFK}.
 
Since one finds  
\be
\label{F1}
F(R_0,c)={kl R_0^2 \over 4\pi G},
\ee
 using (\ref{FRc}) and (\ref{min}), 
Eq.(\ref{slbr2}) has a solution if 
\be
\label{F2}
{kl R_0^2 \over 4\pi G}\leq -8b'\ .
\ee
That puts again some bounds to the dilaton value.
When $|c|$ is small,  using (\ref{min}), one obtains 
\be
\label{F3}
R_0^4\sim {2c^2 \over 3k^2 l^2}\ ,\quad 
F(R_0,c)\sim {1 \over 4\pi G}{|c| \over \sqrt{3}}\ .
\ee
Therefore Eq.(\ref{F2}) is satisfied for small $|c|$. 
On the other hand, when $c$ is large, we get 
\be
\label{F4}
R_0^6\sim {c^2 \over 4k}\ ,\quad 
F(R_0,c)\sim {\left(k |c| \right)^{2 \over 3} \over 
4^{4 \over 3}\pi G}\ .
\ee
Therefore Eq.(\ref{F2}) is not always satisfied and 
we have no solution for $R$ in (\ref{slbr2}) for very large $|c|$. 
Then the existence of the inflationary Universe gives a restriction 
on the value of $c$, which characterizes the behavior of the 
dilaton. 

We now consider the $k<0$ case. When $c=0$, there is no solution for 
$R$ in (\ref{slbr2}). Let us define another function $G(R,c)$ 
as follows:
\be
\label{G1}
G(R,c)\equiv 1 + {l^2 c^2 \over 24 R^8} + {kl^2 \over 3R^2}\ .
\ee
Since $G(R,c)$ appears in the root of $F(R,c)$ in (\ref{FRc}), 
$G(R,c)$ must be positive. Then
\be
\label{G2}
{\partial G(R,c) \over \partial R}=-{l^2 c^2 \over 3R^9}
- {2kl^2 \over 3R^3}\ ,
\ee
$G(R,c)$ has a minimum 
\be
\label{G3}
1+{kl^2 \over 4}\left(-{2k \over c^2}\right)^{1 \over 3},
\ee
when 
\be
\label{G4}
R^6 = -{c^2 \over 2k}\ .
\ee
Therefore if 
\be
\label{G5}
c^2\geq {k^4 l^6 \over 32}\ ,
\ee
$F(R,c)$ is real for any positive value of $R$. Since 
\be
\label{G6}
F(0,c)={|c| \over \pi G \sqrt{24}},
\ee
and when $R\rightarrow \infty$
\be
\label{G7}
F(R,c)\rightarrow {kl R^2 \over 6\pi G}<0\ ,
\ee
there is a solution $R$ in (\ref{slbr2}) if 
\be
\label{G8}
{|c| \over \pi G \sqrt{24}} > -8b'\ .
\ee
If we Wick-rotate the solution corresponding to hyperboloid, 
we obtain a 4 dimensional AdS space, whose metric is given by
\be
\label{Huni}
ds^2_{{\rm AdS}_4}
= dz^2 + \e^{{2z \over R}}\left(-dt^2 + dx^2 + dy^2\right)\ .
\ee
Eq.(\ref{G8}) tells that there is such kind of solution due 
to the quantum effect if the parameter $c$ characterizing the 
behavior of the dilaton is large enough. 
Thus we demonstrated that due to the dilaton presence there is the 
possibility of quantum creation of a 4d hyperbolic wall Universe.
Again, some bounds to the dilaton appear. 
It is remarkable that hyperbolic brane-world occurs even for 
usual matter content due to the dilaton. One can compare with the 
case in ref.\cite{inf} where a hyperbolic 4d wall could be realized 
only for higher derivative conformal scalar. 

In summary, in this section for constant bulk potential, we presented the 
nice 
realization of quantum creation of 4d de Sitter or 4d hyperbolic brane 
Universes living in 5d AdS space. The quantum dynamical 
determination of dilaton value is also remarkable.

\section{\label{II} Non-constant bulk potentials}

We now consider the case that the dilaton field $\phi$ 
has a non-trivial potential:
\be
\label{V}
{12 \over l^2}\ \rightarrow \ 
V(\phi)={12 \over l^2}+\Phi(\phi)\ .
\ee
The surface counter terms when the dilaton field $\phi$ 
has a non-trivial potential are given in \cite{NOO}:
\bea
\label{S1dil}
S^{(2)}&=&S^\phi_1 + S^\phi_2, \nn
S^\phi_1 &=& -{1 \over 16\pi G}\int d^4 \sqrt{\gfr}\left(
{6 \over l} + {l \over 4}\Phi(\phi)\right), \nn
S^\phi_2 &=& -{1 \over 16\pi G}\int d^4 \left\{\sqrt{\gfr}\left(
{l \over 2}R_{(4)} - {l \over 2}\Phi(\phi) \right.\right. \nn
&&\left.\left.
 - {l \over 4}\nabla_{(4)}\phi\cdot \nabla_{(4)}\phi\right)
 - {l^2 \over 8}n^\mu\partial\left(\sqrt{\gfr}\Phi(\phi)\right)
\right\}\ .
\eea
Following the argument in \cite{HHR}, if one replaces 
${12 \over l^2}$ in (\ref{SEHi}) and $S_1$ in (\ref{Stotal}) 
with $V(\phi)$ in (\ref{V}) and  $S^\phi_1$ in (\ref{S1dil}), 
we obtain the gravity on the brane induced by $S^\phi_2$. 
Then if we assume the metric in the following form
\be
\label{DP1}
ds^2=f(y)dy^2 + y\sum_{i,j=0}^3\hat g_{ij}(x^k)dx^i dx^j, 
\ee
as in (\ref{curv2}) and $\phi$ depends only on $y$: $\phi=\phi(y)$, 
we obtain the following equations of motion in the bulk:
\bea
\label{DP2}
0&=&{3 \over 2y^2} - {2kf \over y} - {1 \over 4}\left(
{d\phi \over dy}\right)^2 
- \left({6 \over l^2} + {1 \over 2}\Phi(\phi)\right)f, \\ 
\label{DP3}
0&=&{d \over dy}\left({y^2 \over \sqrt{f}}{d\phi \over dy}\right)
+ \Phi'(\phi)y^2 \sqrt{f}\ .
\eea
On the other hand, on the brane, we obtain the following 
equations instead of (\ref{eq2}) and (\ref{eq2p}):
\bea
\label{eq2b}
0&=&{48 l^4 \over 16\pi G}\left(\partial_z A - {1 \over l}
 - {l \over 24}\Phi(\phi)\right)\e^{4A}
+b'\left(4\partial_\sigma^4 A - 16 \partial_\sigma^2 A\right) \nn
&& - 4(b+b')\left(\partial_\sigma^4 A + 2 \partial_\sigma^2 A 
 - 6 (\partial_\sigma A)^2\partial_\sigma^2 A \right) \nn
&& + 2C\left(\partial_\sigma^4 \phi
 - 4 \partial_\sigma^2 \phi \right), \\
\label{eq2pb}
0&=&-{l^4 \over 8\pi G}\e^{4A}\partial_z\phi
 -{l^5 \over 32\pi G}\e^{4A}\Phi'(\phi) \nn
&& + C\left\{A\left(\partial_\sigma^4 \phi
 - 4 \partial_\sigma^2 \phi \right) 
+ \partial_\sigma^4 (A\phi)
 - 4 \partial_\sigma^2 (A\phi) \right\}\ .
\eea
In (\ref{eq2b}) and (\ref{eq2pb}), one assumes the form of 
the metric as in (\ref{metric1}) instead of (\ref{DP1})  using 
the change of the coordinate: $dz=\sqrt{f}dy$ and equations 
similar to (\ref{smetric}), (\ref{emetric}), and (\ref{hmetric}) 
by choosing $l^2\e^{2\hat A(z,k)}=y(z)$. 

Using (\ref{DP2}) and (\ref{DP3}), we can delete $f$ from the 
equations and we obtain an equation that contains only the 
dilaton field $\phi$:
\bea
\label{DP4}
0&=&\left\{ {5k \over 2} - {k \over 4}y^2
\left({d\phi \over dy}\right)^2 + \left({3 \over 2}y 
 - {y^3 \over 6}\left({d\phi \over dy}\right)^2 \right)
\left({6 \over l^2} + {1 \over 2}\Phi(\phi)\right)\right\}
{d\phi \over dy} \nn
&& + {y^2 \over 2}\left({2k \over y} + {6 \over l^2} 
+ {1 \over 2}\Phi(\phi)\right){d^2\phi \over dy^2}
+ \left({3 \over 4} - {y^2 \over 8}
\left({d\phi \over dy}\right)^2 \right)\Phi'(\phi)\ .
\eea

First we consider a solvable case where
\be
\label{ts1}
{6 \over l^2} + {1 \over 2}\Phi(\phi)= - {2k \over y}\ .
\ee
The explicit form, or $\phi$ dependence, of $\Phi(\phi)$ can 
be determined after solving the equations of motion. 
Then since
\be
\label{ts2}
\Phi'(\phi){d\phi \over dy}={4k \over y^2},
\ee
from (\ref{ts1}), Eq.(\ref{DP4}) can be rewritten as follows:
\be
\label{ts3}
0=\left\{\left({d\phi \over dy}\right)^2 
- {6 \over y^2}\right\}^2\ .
\ee
Note that the $k$-dependence disappears in (\ref{ts3}). 
The solution of (\ref{ts3}) is trivially given by
\be
\label{ts4}
\phi =\pm \sqrt{6} \ln (m^2 y)\ .
\ee
Here $m^2$ is a constant of the integration.\footnote{It is interesting 
that from AdS/CFT point of view the exponent of above dilaton corresponds 
to running gauge coupling which has a power behavior in terms of the energy 
parameter $y$. This gauge coupling corresponds to a boundary QFT with
(broken) supersymmetry.}
 Then from 
(\ref{ts1}) and (\ref{ts2}), we can find the explicit form 
of $\Phi(\phi)$:
\be
\label{ts5}
\Phi(\phi)= - {12 \over l^2}
 - 4km^2\e^{\mp {\phi \over \sqrt{6}}}\ .
\ee
Note that exponential potentials of the above type often appear as 
the result of spherical reduction in M-theory or string theory, 
see discussion in ref.\cite{cvetic}.
One can also find that Eq.(\ref{DP2}) is trivially satisfied. 
Integrating (\ref{DP3}), we obtain
\be
\label{ts6}
f={1 \over -{2ky \over 9} + {f_0 \over y^2}}\ .
\ee
Here $f_0$ is a constant of the integration and $f_0$ should be 
positive in order that $f$ is positive for large $y$. 
There is a (curvature) singularity at $y=0$. 
One should also note that when $k>0$, the horizon appears  at
\be
\label{ts7}
y^3 = y_0^3\equiv {9f_0 \over 2k}\ ,
\ee
and we find 
\be
\label{ts7b}
y\leq y_0\ .
\ee
Then since
\be
\label{ts8}
\partial_z A = {1 \over 2}\partial_z\left(\ln y\right) 
= {1 \over 2y}{dy \over dz}={1 \over 2y \sqrt{f(y)}} \ ,
\ee
Eqs.(\ref{eq2b}) and (\ref{eq2pb}) have the following forms: 
\bea
\label{ts9}
0&=&{1 \over \pi G}\left({1 \over 2y_0}\sqrt{
{f_0 \over y_0^2} - {2ky_0 \over 9}} - {1 \over 2l}
+ {kl \over 3y_0}\right)y_0^2 + 8b' \nn
&=&{1 \over \pi G}\left({1 \over 2R^2}\sqrt{
{f_0 \over R^4} - {2kR^2 \over 9}} - {1 \over 2l}
+ {kl \over 3R^2}\right)R^4 + 8b', \\
\label{ts10}
0&=& - {y_0\sqrt{6} \over 8\pi G}\sqrt{
{f_0 \over y_0^2} - {2ky_0 \over 9}} - {kly_0\sqrt{6} 
\over 2\pi G } + 6C\phi_0 \nn
&=& - {R^2\sqrt{6} \over 8\pi G}\sqrt{
{f_0 \over R^4} - {2kR^2 \over 9}} - {klR^2\sqrt{6} 
\over 48\pi G } + 6C\phi_0 \ .
\eea
Eq.(\ref{ts10}) gives a value of the dilaton on the 
brane:
\be
\label{ts11}
\phi_0={1 \over C\sqrt{6}}{R^2 \over 8\pi G}
\left(\sqrt{{f_0 \over R^4} - {2kR^2 \over 9}} + {kl \over 6}
\right)\ .
\ee
When $k>0$, Eq.(\ref{ts9}) does not have a solution for large 
$R$ since there is an upper bound $R\leq \sqrt{y_0}$ coming 
from (\ref{ts7b}). Even for $k\leq 0$, there is no solution for 
large $R$ in case of ${\cal N}=4$ Yang-Mills theory ($b'<0$) 
since Eq.(\ref{ts9}) behaves for large $R$
\be
\label{ts12}
0\sim -{1 \over 2l\pi G}R^4 + 8b'  \ .
\ee
On the other hand, if one assumes $R$ is small, Eq.(\ref{ts9}) 
has the following form:
\be
\label{tssR1}
0= {1 \over \pi G}\left({\sqrt{f_0} \over 2R^4} 
+ {kl \over 3R^2}\right)R^4 + 8b' + {\cal O}(R^2)\ ,
\ee
which can be solved with respect to $R$:
\be
\label{tssR2}
R^2 = - {3 \over kl}\left({\sqrt{f_0} \over 2} 
+ 8\pi G b'\right)\ .
\ee
Then there is a solution for $k<0$ ($k>0$) if 
\be
\label{tssR3}
f_0 > 128\pi^2 G^2 {b'}^2 \quad 
\left(f_0 < 128\pi^2 G^2 {b'}^2\right)\ .
\ee
Hence, the results are similar to those in the previous section but
in the presence of non-trivial bulk potential.

One can also consider the case of no quantum corrections, i.e. 
 $W$ vanishes.  Putting $C=b'=0$, we obtain from (\ref{ts9}) 
and (\ref{ts10})
\bea
\label{tsc1}
0&=&{1 \over 2R^2}\sqrt{
{f_0 \over R^4} - {2kR^2 \over 9}} - {1 \over 2l}
+ {kl \over 3R^2}, \\
\label{tsc2}
0&=&\sqrt{{f_0 \over R^4} - {2kR^2 \over 9}} + {kl \over 6}\ .
\eea
Eq.(\ref{tsc2}) tells that $k\leq 0$ but by combining (\ref{tsc1}) 
and (\ref{tsc2}), we find $R^2={kl^2 \over 2}$. Then there is not 
consistent solution. 

Note, however, that the quantum equation (65) for $R$ has the solution for 
conformally invariant higher derivative scalar whose contribution to $b'$
is positive: $b=-8/120(4 \pi)^2, b'=28/360(4 \pi)^2$. In a similar 
way one can analyze other types of dilatonic potentials (numerically 
or using some perturbative technique) which lead to (singular) 5d AdS space 
with 4d constant curvature wall(s).

Let us discuss other examples in attempt to construct 
non-singular brane-world with inflationary brane induced by
quantum effects.
As the singularity usually appears at $y=0$, we investigate 
the behavior of (\ref{DP4}) when $y\sim 0$. 
Here we only consider the case $k>0$. 
First one assumes that there is no singularity. Then $\phi$, 
${d\phi \over dy}$, and ${d^2 \phi \over dy^2}$ would be 
finite and we can assume
\be
\label{DP5}
\phi\rightarrow \phi_1\ (\mbox{constant})\ \mbox{when} \ 
y\rightarrow 0\ .
\ee
It is supposed the spacetime becomes asymptotically AdS, which is  
presumbly the unique choice to avoid the singularity and to localize  
gravity on the brane \cite{CEGH}. The condition to get 
asymptotically AdS requires 
\be
\label{DP6}
\Phi'(\phi_1)=0,\ 
\ee
and one assumes 
\be
\label{DP7}
\Phi'(\phi)\sim \beta \phi_2^\alpha\ (\alpha>0), \quad 
\phi_2\equiv \phi - \phi_1\ .
\ee
Then from (\ref{DP4}), one gets
\be
\label{DP8}
0\sim {5k \over 2}{d\phi_2 \over dy} 
+ ky {d^2\phi_2 \over dy^2} + {3 \over 4}\beta \phi_2^\alpha\ .
\ee
If we also assume $\phi_2$ behaves as 
\be
\label{DP9}
\phi_2\sim \tilde b y^a\ (a>0)\ ,
\ee
one  obtains 
\bea
\label{DP10}
\alpha&=&1 - {1 \over a} \\
\label{DP11}
\beta&=& - {4k \over 3}{\tilde b}^{1 \over a}a\left(a 
+ {3 \over 2}\right)\ .
\eea
Eq.(\ref{DP10}) requires $0<\alpha<1$ and/or $a>1$ and 
Eq.(\ref{DP11}) tells that $\beta$ cannot vanish and 
$\tilde b$ should be positive, which tells that $\phi$ increases when 
$y\sim 0$. 

If we assume ${d\phi \over dy}=0$ at $y=y_1>0$ in (\ref{DP4}), 
we obtain
\be
\label{DP12}
0={y_1^2 \over 2}\left({2k \over y_1} + {6 \over l^2} 
+ {1 \over 2}\Phi(\phi(y_1))\right){d^2\phi \over dy^2}
+ {3 \over 4} \Phi'(\phi(y_1))\ .
\ee
In case that $V(\phi)={12 \over l^2} + \Phi(\phi)>0$, 
${d^2\phi \over dy^2}>0$ $\left({d^2\phi \over dy^2}<0\right)$ 
if $\Phi'(\phi)<0$ $\left(\Phi'(\phi)>0\right)$. 
Since $\phi$ increases when $y\sim 0$, $\phi$ increases 
monotonically if $V(\phi)>0$ and $\Phi'(\phi)<0$. 

One also finds that when $y$ is large, Eq.(\ref{DP4}) does not 
depend on $k$:
\bea
\label{DP13}
0&=&\left({3 \over 2}y 
 - {y^3 \over 6}\left({d\phi \over dy}\right)^2 \right)
\left({6 \over l^2} + {1 \over 2}\Phi(\phi)\right)
{d\phi \over dy} \nn
&& + {y^2 \over 2}\left( {6 \over l^2} 
+ {1 \over 2}\Phi(\phi)\right){d^2\phi \over dy^2}
+ \left({3 \over 4} - {y^2 \over 8}
\left({d\phi \over dy}\right)^2 \right)\Phi'(\phi)\ .
\eea

Let us consider the following example as a toy model:
\be
\label{DP14}
l^2 \Phi(\phi) = - {4 \over 3}\phi^{3 \over 2} 
+ {3 \over 4}\phi^4 - {1 \over 8}\phi^8  
+ {17 \over 24}\ .
\ee
Since 
\be
\label{DP15}
l^2 \Phi'(\phi)=-2\phi^{1 \over 2} + 3\phi^3 - \phi^7 \ ,
\ee
by comparing (\ref{DP15}) with (\ref{DP9}), one finds
\be
\label{DP16}
\phi_1=0\ ,\quad \alpha={1 \over 2}\ ,\quad \beta=-2\ ,
\quad a=2\ ,\quad \tilde b={1 \over 196}\ .
\ee
Eq.(\ref{DP15}) also tells $\Phi'(\phi)=0$ when $\phi=0$ or $1$ 
and $\Phi'(\phi)<0$ when $0<\phi<1$. 
Then if $\phi\rightarrow 0$ when $y\rightarrow 0$, 
we can naively expect $\phi\rightarrow 1$ when 
$y\rightarrow +\infty$. This naive expectation can be confirmed 
by the numerical calculation, which is given in Figs.\ref{Fig1} 
and {\ref{Fig2}. 
\begin{figure}
\epsffile{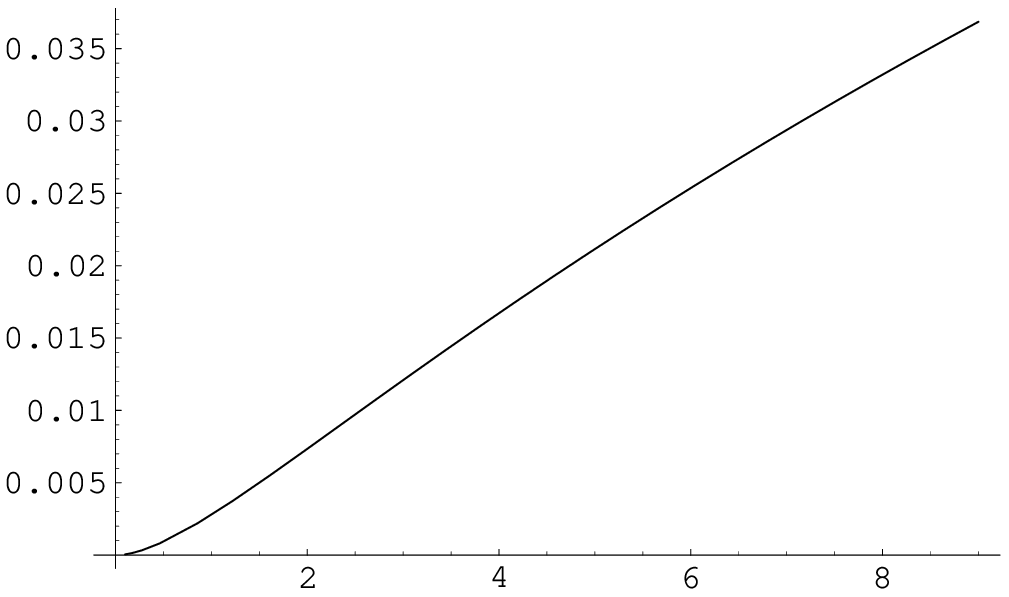}
\caption{The behavior of $\phi$ (vertical axis) versus 
${y \over l^2}$ (horizontal axis) when $y$ is small. }
\label{Fig1}
\end{figure}
\begin{figure}
\epsffile{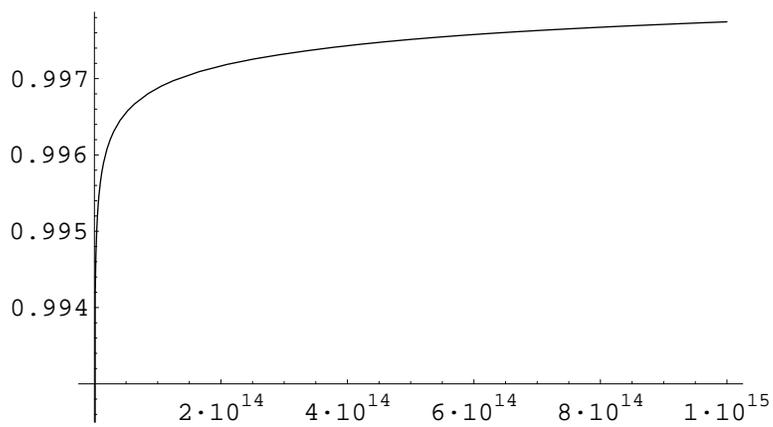}
\caption{The behavior of $\phi$ (vertical axis) versus 
${y \over l^2}$ (horizontal axis) when $y$ is large. }
\label{Fig2}
\end{figure}
In Fig.\ref{Fig1}, the behavior of $\phi$ when $y$ is small is 
given and in Fig.\ref{Fig2}, the behavior of $\phi$ when $y$ is 
large is drawn. From Fig.\ref{Fig2}, one can find that $\phi$ goes 
to unity when $y$ is large. Then there is not any (curvature) 
singularity and the gravity on the brane can be localized. 
If we assume 
\be
\label{DP17}
1-\phi \sim \eta y^\xi\ ,\quad \xi<0\ ,\quad \mbox{($\eta$ and 
$\xi$ are constant).}
\ee
the numerical calculation in Fig.\ref{Fig2} tells
\be
\label{DP19}
\xi=-0.2\ ,\quad \eta=1\ .
\ee
Then from (\ref{DP2}), we find the behavior of $f(y)$ when $y$ 
is large:
\be
\label{DP20}
f\sim {l^2 \over 4y^2}\left\{1 - {\xi^2 \over 6}\eta^2 
\left({y \over l^2}\right)^{2\xi} 
+ \cdots \right\}\ .
\ee
When $y_0$ is large, Eqs.(\ref{eq2b}) and (\ref{eq2pb}) have 
the following forms:
\bea
\label{DP21}
0&\sim&{l^3\xi^2\eta^2 \over 12\pi G}
\left({y_0 \over l^2}\right)^{2\xi+2} + 8b' \nn
&=&{l^3\xi^2\eta^2 \over 12\pi G}
\left({R \over l}\right)^{4\xi+4} + 8b', \\
\label{DP22}
0&\sim&{l^3\eta \xi \over 4\pi G }\left({y_0 \over l^2}\right)^{\xi +2} 
+ 6C\phi_0 \nn
&=&{l^3\eta \xi \over 4\pi G }\left({R \over l}\right)^{2\xi +4} 
+ 6C\phi_0 \ .
\eea
Eqs.(\ref{DP21}) and (\ref{DP22}) can be solved with 
respect to $R$ and $\phi_0$, respectively:
\bea
\label{DP22b}
R&\sim&l\left(-{96\pi G b' \over l^3\xi^2 \eta^2}
\right)^{1 \over 4+4\xi}, \\ 
\label{DP23}
\phi_0&\sim&\left(-{4b' \over C}\right)\left(\eta\xi
\right)^{-{1 \over 1+\xi}}\left(- {l^3 \over 96\pi G b'}
\right)^{\xi \over 2+2\xi}\ .
\eea
Since $-b'={C \over 4}
={N^2 -1 \over 4(4\pi)^2}$ from (\ref{bbC}) for ${\cal N}=4$ 
$SU(N)$ Yang-Mills theory and ${G \over l^3}={\pi \over 2N^2}$, 
Eqs.(\ref{DP22b}) and (\ref{DP23}) tell that 
\bea
\label{DP24}
R&\sim& l \left({3 \over 4\eta^2\xi^2}\right)^{1 \over 4 + 4\xi}
=2.5\cdots, \\
\label{DP25}
\phi_0&\sim& \left(\eta\xi\right)^{-{1 \over 1+\xi}}
\left({4 \over 3}\right)^{\xi \over 2 + 2\xi}=0.13\cdots \ .
\eea
Here we also used (\ref{DP19}). 
Since $R$ is not so large, the large $y$ or $R$ approximation 
converges slowly. Since $0<\phi_0<1$, however, there is no 
apparent conflict. 
Eq.(\ref{DP24}) shows that the brane does not lie in the 
asymptotically AdS region when $y$ is large. Anyway it  
suggests that there is a solution where 
the brane corresponds to  $S_4$, which gives the de Sitter space after 
transition to lorentzian signature.

For comparison, one 
can consider the classical case where $W$ vanishes. Then 
(\ref{eq2b}) and (\ref{eq2pb}) have the following form:
\bea
\label{DPc1}
0&=&{1 \over 2y\sqrt{f(y)}} - {1 \over l} 
 - {l \over 24}\Phi(\phi), \\
\label{DPc2}
0&=& {1 \over \sqrt{f(y)}}{d\phi \over dy}
 - {l \over 4}\Phi'(\phi) \ .
\eea
Here it is used
\be
\label{DPc3}
\partial_z A = {1 \over 2y\sqrt{f(y)}}\ ,\quad 
{d \over dz}={1 \over \sqrt{f(y)}}{d \over dy}\ .
\ee
Since we have 
\be
\label{DPc4}
f(y)={{3 \over 2y^2} - {1 \over 4}\left(
{d\phi \over dy}\right)^2 \over  {2k \over y} 
+ {6 \over l^2} + {1 \over 2}\Phi(\phi)},
\ee 
from (\ref{DP2}), one can delete $f$ and ${d\phi \over dy}$ in 
(\ref{DPc1}), (\ref{DPc2}), (\ref{DPc4}) and obtain 
\be
\label{DPc5}
{2k \over y_0}=l^2\left( -{3 \over 8}\Phi'(\phi) 
+ {1 \over 96}\Phi^2\right)\ .
\ee
For the potential (\ref{DP14}), the l.h.s. in (\ref{DPc5}) 
is negative when $1>\phi>\phi_0\sim 0.000144\cdots$, which is 
almost all the allowed region ($1>\phi>0$) in the solution for 
the potential in (\ref{DP14}). Therefore there is no  
classical solution for the $k>0$ case. Then the brane solution 
corresponding to 4 dimensional sphere or de Sitter space cannot 
exist without quantum correction coming from $W$. 

Thus, using fine-tuned dilatonic potential in AdS dilatonic gravity 
we presented non-singular asymptotically AdS bulk space with de Sitter 
brane living on the boundary. The dilatonic de Sitter brane is induced by 
quantum effects of the CFT on the wall. As one can see, gravity trapping 
occurs. 
The values of brane radius and of dilaton are dynamically determined.

\section{\label{III}Not exactly conformal brane quantum matter} 

In this section, we consider the case that the matter on the 
brane is 
not the exact CFT like super Yang-Mills theory but some exactly 
non-conformal theory like QED or QCD. Of course, such a theory 
is classically a conformally invariant one. As an explicit example in
order to be able to apply large N-expansion we suppose that dominant 
contribution is due to $N$ massless Majorana spinors coupled with the 
dilaton, whose action is given by
\be
\label{SP1}
S=\int \sqrt{\gfr} \e^{a\phi}\sum_{i=1}^N\bar\Psi_i \gamma^\mu
D_\mu\Psi_i\ .
\ee
The case of minimal spinor coupling corresponds to the choice $a=0$.
Then the trace anomaly induced action $W$ corresponding to 
(\ref{W}) has the following form \cite{NOZ}:
\bea
\label{W2}
W&=& b \int d^4x \sqrt{\widetilde g}\widetilde F A_1 \nn
&& + b' \int d^4x\left\{A_1 \left[2{\widetilde\Box}^2 
+\widetilde R_{\mu\nu}\widetilde\nabla_\mu\widetilde\nabla_\nu 
 - {4 \over 3}\widetilde R \widetilde\Box^2 
+ {2 \over 3}(\widetilde\nabla^\mu \widetilde R)\widetilde\nabla_\mu
\right]A_1 \right. \nn
&& \left. 
+ \left(\widetilde G - {2 \over 3}\widetilde\Box \widetilde R
\right)A_1 \right\} \\
&& -{1 \over 12}\left\{b''+ {2 \over 3}(b + b')\right\}
\int d^4x \left[ \widetilde R - 6\widetilde\Box A_1 
 - 6 (\widetilde\nabla_\mu A_1)(\widetilde \nabla^\mu A_1)
\right]^2 \ .\nonumber
\eea
Here 
\be
\label{SP2}
A_1=A+{a\phi \over 3},
\ee
and 
\be
\label{SP3}
b={3N \over 60(4\pi)^2}\ ,\quad b'=-{11 N \over 360 (4\pi)^2}\ .
\ee
We also choose $b''=0$ as it may be changed by finite renormalization 
of classical gravitational action.

First one considers a constant potential  
($\Phi(\phi)=0$). Then the behavior of the solution in the bulk 
do not change with respect to those in Section \ref{I}. On the brane, we 
obtain the following equations corresponding to 
(\ref{eq2}) and (\ref{eq2p}):
\bea
\label{eq2c}
0&=&{48 l^4 \over 16\pi G}\left(\partial_z A - {1 \over l}
\right)\e^{4A}
+b'\left(4\partial_\sigma^4 A_1 - 16 \partial_\sigma^2 A_1
\right) \nn
&& - 4(b+b')\left(\partial_\sigma^4 A_1 + 2 \partial_\sigma^2 A_1 
 - 6 (\partial_\sigma A_1)^2\partial_\sigma^2 A_1 \right), \\
\label{eq2pc}
0&=&-{l^4 \over 8\pi G}\e^{4A}\partial_z\phi
+ {4 \over 3}ab' \left(4\partial_\sigma^4 A_1 
- 16 \partial_\sigma^2 A_1\right) \ .
\eea
Then one gets  
\bea
\label{SP4}
0&=&{1 \over \pi G l}\left\{\sqrt{1 + {kl^2 \over 3R^2} 
+ {l^2 c^2 \over 24 R^8}} -1 \right\}R^4 + 8 b', \\
\label{SP5}
0&=& - {c \over 8\pi G} + 32 ab' \ .
\eea
Note that for minimal spinor coupling the second equation does not have 
a solution.
Eq.(\ref{SP4}) is identical with (\ref{slbr2}). 
Eq.(\ref{SP5}) can be solved with respect to $c$:
\be
\label{SP5b}
c=32\times 8\pi G a b',
\ee
but the boundary value $\phi_0$ of $\phi$ becomes a free 
parameter.  Hence, for constant bulk potential there is again the 
possibility of quantum creation of a 4d de Sitter or a 4d hyperbolic 
brane living in 5d AdS bulk space. This occurs even for not exactly 
conformal 
invariant quantum brane matter. The details of this scenario are similar to 
those in section 2.

When there is a non-trivial potential corresponding to 
(\ref{ts1}) or (\ref{ts5}), Eq.(\ref{ts9}) is not changed 
but Eq.(\ref{ts10}) is changed into
\be
\label{ts10c}
0= - {R^2\sqrt{6} \over 8\pi G}\sqrt{
{f_0 \over R^4} - {2kR^2 \over 9}} - {klR^2\sqrt{6} 
\over 2\pi G } + 32 ab' \ .
\ee

For the potential (\ref{DP14}), Eq.(\ref{DP21}) is not changed 
again and instead of (\ref{DP22}), we obtain
\be
\label{DP22bc}
0={l^3\eta\xi \over 4\pi G l} \left({R \over l}\right)^{2\xi +4} 
 + 32 ab' \ .
\ee
Eqs.(\ref{ts10c}) and (\ref{DP22bc}) define the 
parameter $a$, which characterizes the dilaton coupling 
in (\ref{SP1}). 
Since the equations for $R$ are identical with (\ref{ts9}) and 
(\ref{DP21}), the expression of the radius is not changed.  
Then for the potential (\ref{DP14}), we have 
\be
\label{LLL1}
R\sim l\left(-{96\pi G b' \over l^3\xi^2 \eta^2}
\right)^{1 \over 4+4\xi} \ ,
\ee
if $R$ is large. 
In this case, however, the value of $b'$ in (\ref{SP3})is 
different from that of (\ref{bbC}) for ${\cal N}=4$ 
$SU(N)$ Yang-Mills theory and we do not know the value 
for ${l^3 \over G}$ (it may be considered as free parameter). Then the 
value of $R$ itself will be changed from the one in the previous section. 

Hence, we have shown that in case of quantum brane matter different 
from super Yang-Mills theory still there arises (non)-singular 
brane-worlds for various dilatonic potentials in d5 AdS dilatonic gravity.
As in the previous section gravity is trapped. The brane represents a 
constant  curvature space which may be considered as an inflationary phase of 
our observable Universe.

\section{Discussion}

In summary, the role of brane quantum matter effects in 
the realization of de Sitter or AdS dilatonic branes living 
in d5 (asymptotically) AdS space is carefully investigated. 
(We are working with d5 dilatonic gravity).
The explicit examples of such dilatonic brane-world inflation are 
presented for constant bulk dilatonic potentials 
as well as for non-constant bulk potentials. Dilaton gives extra 
contributions to the effective tension of the domain wall and it may 
be determined dynamically from bulk/boundary equations of motion. 
The main part of discussion  has dealing with maximally SUSY Yang-Mills 
theory (exact CFT) living on the brane.
However, in section 4 we demonstrated that qualitatively 
the same results may be obtained 
when not exactly conformal quantum matter (like 
classically conformally invariant theory of dilaton coupled spinors) 
lives on the brane. An explicit example of toy (fine-tuned) 
dilatonic potential is presented 
for which the following results are obtained from the bulk/boundary 
equations of motion:
1. Non-singular asymptotically AdS space is the bulk space.
2. The brane is described by de Sitter space (inflation) induced by
brane matter quantum effects.
3. The localization of gravity on the brane occurs.
The price to avoid the bulk naked singularity is the fine-tuning 
of dilatonic potential and dynamical determination (actually,
also a kind of fine-tuning) of dilaton and radius of de Sitter brane.
Note also that in the same fashion as in ref.\cite{HHR} one can 
show that the brane CFT strongly suppresses the metric perturbations 
(especially, on small scales).

One can easily generalize the results of this work in different 
directions. For example, taking into account that it is not 
easy to find new dilatonic 
bulk solutions like asymptotically AdS space presented in this work
one can think about changes in the structure of the boundary manifold.
One possibility is in the consideration of a Kantowski-Sachs brane 
Universe. Another important question is related with the study 
of cosmological perturbations around the founded backgrounds and of 
details of late-time inflation and exit from inflationary phase 
in brane-world cosmology (eventual decay of de Sitter brane to FRW 
brane).  
It would be also interesting to study more examples of dilatonic 
potentials within the action and brane-world structure under 
consideration. Clearly, this can be done  numerically or 
using some perturbative expansion of the potential.

\ 

\noindent
{\bf Acknowledgements.} 
The work by SDO has been supported in part by CONACyT (CP, ref.990356) 
and in part by RFBR. This research has been also supported in part by
CONACyT grant 28454E.


\section*{Appendix}

AdS$_5$/CFT$_4$ correspondence tells us that the effective 
action $W_{\rm CFT}$ of CFT in 4 dimensions is given by the path 
integral of the supergravity in 5 dimensional AdS space:
\bea
\label{A1}
\e^{-W_{\rm CFT}}&=&\int [dg][d\varphi]\e^{-S_{\rm grav}}, \\
S_{\rm grav}&=&\SEH + \SGH + S_1 + S_2 + \cdots, \nn
\SEH&=&{1 \over 16\pi G}\int d^5 x \sqrt{\gfv}\left(R_{(5)} 
 + {12 \over l^2} + \cdots \right), \nn
\SGH&=&{1 \over 8\pi G}\int d^4 x \sqrt{\gfr}\nabla_\mu n^\mu, \nn
S_1&=& -{1 \over 8\pi G l}\int d^4 x \sqrt{\gfr}\left({3 \over l}
+ \cdots \right), \nn
S_2&=& -{1 \over 16\pi G l}\int d^4 x \sqrt{\gfr}\left(
{1 \over 2}R_{(4)} + \cdots \right), \nn
&& \cdots \ . \nonumber
\eea
Here $\varphi$ expresses the (matter) fields besides the graviton. 
$\SEH$ corresponds to the Einstein-Hilbert action and 
$\SGH$ to the Gibbons-Hawking surface counter term and 
$n^\mu$ is the unit vector normal to the boundary. 
$S_1$, $S_2$, $\cdots$ correspond to the surface counter terms, 
which cancell the divergences when the boundary in AdS$_5$ goes to 
the infinity. 

In \cite{HHR}, two 5 dimensional balls $B_5^{(1,2)}$ are 
glued on the boundary, which is 4 dimensional sphere $S_4$. Instead 
of $S_{\rm grav}$, if one considers the following action $S$
\be
\label{A2}
S=\SEH + \SGH + 2S_1=S_{\rm grav} + S_1 - S_2 - \cdots,
\ee
for two balls, using (\ref{A1}), one gets the following 
boundary theory in terms of the partition function \cite{HHR}:
\bea
\label{A3}
&& \int_{B_5^{(1)} + B_5^{(1)} +S_4} [dg][d\varphi]\e^{-S} \nn
&=& \left(\int_{B_5} [dg][d\varphi]\e^{-\SEH - \SGH - S_1} 
\right)^2 \nn
&=&\e^{2S_2 + \cdots}\left(\int_{B_5} [dg][d\varphi]
\e^{-S_{\rm grav}} \right)^2 \nn
&=&\e^{-2W_{\rm CFT}+2S_2 + \cdots}\ .
\eea
Since $S_2$ can be regarded as the Einstein-Hilbert action on 
the boundary, which is $S_4$ in the present case, the gravity 
on the boundary becomes dynamical. The 4 dimensional gravity 
is nothing but the gravity localized on the brane in the 
Randall-Sundrum model \cite{RS}. 

For ${\cal N}=4$ $SU(N)$ Yang-Mills theory, the AdS/CFT dual is 
given by identifying
\be
\label{AdSCFT}
l=g_{\rm YM}^{1 \over 2}N^{1 \over 4}l_s\ ,
\quad {l^3 \over G}={2N^2 \over \pi}\ .
\ee
Here $g_{\rm YM}$ is the coupling of the Yang-Mills theory and 
$l_s$ is the string length. Then (\ref{A3}) tells that the 
RS model is equivalent to  a CFT (${\cal N}=4$ $SU(N)$ 
Yang-Mills theory) coupled to 4 dimensional gravity including 
some correction coming from the higher order counter terms with 
a Newton constant given by
\be
\label{coupling}
G_4=G/l\ .
\ee

This is an excellent explanation \cite{HHR} to why gravity is trapped on the 
brane.

In case that we include the dilaton field (generally with the 
dilaton potential), the explicit forms of the actions are given 
by 
\bea
\label{Aactions}
\SEH^\phi&=&{1 \over 16\pi G}\int d^5 x \sqrt{\gfv}\left(R_{(5)} 
 -{1 \over 2}\nabla_\mu\phi\nabla^\mu \phi 
 + {12 \over l^2}+\Phi(\phi) \right), \nn
S_1^\phi &=& -{1 \over 16\pi G}\int d^4 \sqrt{\gfr}\left(
{6 \over l} + {l \over 4}\Phi(\phi)\right), \nn
S_2^\phi &=& -{1 \over 16\pi G}\int d^4 \left\{\sqrt{\gfr}\left(
{l \over 2}R_{(4)} - {l \over 2}\Phi(\phi) \right.\right. \nn
&&\left.\left.
 - {l \over 4}\nabla_{(4)}\phi\cdot \nabla_{(4)}\phi\right)
 - {l^2 \over 8}n^\mu\partial_\mu\left(\sqrt{\gfr}\Phi(\phi)\right)
\right\}\ .
\eea
 AdS/CFT tells the effective action $W$ of the boundary 
field theory is given by
\be
\label{A4}
\e^{-W}=\int [dg][d\phi][d\tilde\varphi]\e^{-\SEH^\phi 
 - \SGH - S_1^\phi - S_2^\phi + \cdots } .
\ee
Here $\tilde\varphi$ express the fields besides the graviton 
and dilaton. 
Then if we consider the action
\be
\label{A5}
S=\SEH^\phi + \SGH + 2S_1^\phi, 
\ee
in the two balls, instead of (\ref{A2}), 
we obtain the following 
boundary theory given, instead of (\ref{A3}): 
\bea
\label{A6}
&& \int_{B_5^{(1)} + B_5^{(1)} +S_4} [dg][d\phi]
[d\tilde\varphi]\e^{-S} \nn
&=& \left(\int_{B_5} [dg][d\varphi]
\e^{-\SEH^\phi - \SGH - S_1^\phi} \right)^2 \nn
&=&\e^{-2W+2S_2^\phi + \cdots}\ .
\eea
As $S_2^\phi$ contains the Einstein action, 
there appears the dilatonic gravity localized on the brane. 
We can also choose the trace anomaly induced action as 
the effective action $W$. 

Then again in this case, by using the identifications in 
(\ref{AdSCFT}) and {coupling}, (\ref{A6}) tells that the 
dilatonic RS model is equivalent to  a CFT coupled to 4d 
dilatonic gravity. Such equivalence may be useful
in various explicit calculations.


\begin{thebibliography}{99}
\bibitem{RS} L. Randall and R. Sundrum,
 {\sl Phys.Rev.Lett.} {\bf 83} (1999) 3370, hep-th/9905221;
 {\sl Phys.Rev.Lett.} 
 {\bf 83} (1999)4690, hep-th/9906064. 
\bibitem{CH} A.H. Chamblin and H.S. Reall, hep-th/9903225.
 N. Kaloper, {\sl Phys.Rev.} {\bf D60} (1999) 123506,
hep-th/9905210;
 T. Nihei, {\sl Phys.Lett.} {\bf B465} (1999) 81, hep-th/9905487;
 H. Kim and H. Kim, hep-th/9909053;
 C. Csaki, M. Graesser, C. Kolda and J. Terning,
 hep-ph/9906513;
 J. Cline, C. Crojean and G. Servant, {\sl Phys.Rev.Lett.} {\bf 83}
(1999) 4245, hep-ph/9906523;
 D.J. Chung and K. Freese, {\sl Phys.Rev.} {\bf D61} (2000) 023511,
hep-ph/9906542;
 R. Maartens, D.Wands, B. Bassett and T. Heard, hep-ph/9912464;
 P. Kanti, I. Kogan, K. Olive and M. Pospelov, {\sl Phys.Lett.}
 {\bf B468} (1999) 31, hep-ph/9909481;
 P. Binetruy, C. Deffayet, U. Ellwanger and D. Langlois,
 {\sl Phys.Lett.} {\bf B477} (2000) 285,
 hep-th/9910219;
 S. Mukohyama, T. Shiromizu and K. Maeda, hep-th/9912287;
 J. Garriga and M. Sasaki, hep-th/9912118;
 K. Koyama and J. Soda, hep-th/9912118;
 J. Kim, B. Kyae and H. Min Lee, hep-th/0004005.
\bibitem{HHR} S.W. Hawking, T. Hertog and H.S. Reall, 
hep-th/0003052, {\sl Phys.Rev.} {\bf D62} (2000) 043501.
\bibitem{NOZ} S. Nojiri, S.D. Odintsov and S. Zerbini,
hep-th/0001192, {\sl Phys.Rev.} {\bf D}, to appear;
 S. Nojiri, O. Obregon, S.D. Odintsov
and S. Ogushi, hep-th/0003148, {\sl Phys.Rev.} {\bf D}, to appear.
\bibitem{ADKS} N. Arkani-Hamed, S. Dimopoulos, N. Kaloper and 
R. Sundrum, hep-th/0001197;
S. Kachru, M. Schultz and E. Silverstein, hep-th/0001206.
\bibitem{ADDK}
N. Arkani-Hamed, S. Dimopulos, Dvali and Kaloper, 
: hep-th/9907209, {\sl Phys.Rev.Lett.} {\bf 84} (2000) 586.
\bibitem{Guendelman} E.I.Guendelman,  
{\sl Mod.Phys.Lett.} {\bf A14} (1999) 1397; 
gr-qc/9906025, {\sl Class.Quant.Grav.} {\bf 17} (2000) 361; 
gr-qc/9901067; gr-qc/9901017 {\sl Mod.Phys.Lett.} {\bf A14} 
(1999) 1043. 
\bibitem{Youm} D. Youm, hep-th/0002147;
J. Chen, M. Luty and E. Ponton, hep-th/0003067;
S.P. de Alwis, A.T. Flournoy and N. Irges, hep-th/0004125. 
\bibitem{Pot} K. Behrndt and M. Cvetic, hep-th/9909058,
hep-th/0001159;
A. Chamblin and G. Gibbons, {\sl Phys.Rev.Lett.} {\bf 84} (2000) 1090,
hep-th/9909130;
O. De Wolfe, D. Freedman, S. Gubser and A. Karch, hep-th/9909134.
\bibitem{BOS} I.L. Buchbinder, S.D. Odintsov and I.L. Shapiro,
Effective Action in Quantum Gravity,
IOP Publishing, Bristol and Philadelphia 1992.
\bibitem{LT} H. Liu and A. Tseytlin, {\sl Nucl.Phys.} {\bf B533} 
(1998) 88, hep-th/9804083;
S. Nojiri and S.D. Odintsov, {\sl Phys.Lett.} {\bf B444} (1998) 92.
\bibitem{inf} S. Nojiri and S.D. Odintsov, hep-th/0004097.

\bibitem{SMM} A. Starobinsky, {\sl Phys.Lett.} {\bf B91} (1980) 99;
S.G. Mamaev and V.M. Mostepanenko, {\sl JETP} {\bf 51} (1980) 9.

\bibitem{NOtwo} S. Nojiri and S.D. Odintsov, {\sl Phys.Lett.}
{\bf B449} (1999) 39, hep-th/9812017;
{\sl Phys.Rev.} {\bf D61} (2000) 044014, hep-th/9905200;
\bibitem{NOO}
S. Nojiri, S.D. Odintsov and S. Ogushi, hep-th/0001122. 

\bibitem{GJS} C. G\'omez, B. Janssen and P. Silva, 
hep-th/0002042. 

\bibitem{CEGH} C. Cs\'aki, J. Erlich, C. Crojean and 
T.J. Hollowood, hep-th/0004133.  
\bibitem{Brevik} I. Brevik and S.D. Odintsov, {\sl Phys.Lett.} 
 {\bf B455} (1999) 104, hep-th/9902418;
B. Geyer, S.D. Odintsov and S. Zerbini, {\sl Phys.Lett.} {\bf B460} 
(1999) 58.
\bibitem{cvetic} M. Bremer, M. Duff, H. Lu, C. Pope, and K.S. Stelle,
hep-th/9807051, {\sl Nucl.Phys.} {\bf B543} (1999) 329;
M. Cvetic, H. Lu and C.N. Pope, hep-th/0001002, hep-th/0003286. 
\bibitem{AFK} S. Abel, K. Freese and I. Kogan, hep-th/0005028.

\end{thebibliography}
\end{document}